\documentclass[preprints,article,accept,moreauthors,pdftex]{Definitions/mdpi}

\usepackage{doi}
\usepackage{textcomp}
\usepackage{tikz}
\usetikzlibrary{decorations.pathmorphing}
\usetikzlibrary{decorations.markings}
\usepackage{units}

\firstpage{1} 
\pubvolume{1}
\issuenum{1}
\articlenumber{0}
\pubyear{2021}
\copyrightyear{2021}
\externaleditor{{Academic Editor: Firstname Lastname}} 
\datereceived{25 June 2021} 
\dateaccepted{14 July 2021} 
\datepublished{} 
\hreflink{https://doi.org/} 
\pdfoutput=1

\newcommand{\lsi}{LS~I~+61\textdegree{}303}
\newcommand{\gammaray}{$\gamma$-ray}
\newcommand{\fermilat}{\textit{Fermi}-LAT}
\newcommand{\ltm}{LTM}
\newcommand{\halpha}{H$\alpha$}

\Title{A Precessing Jet Scenario for the Multi-Wavelength Long-Term Modulation of \lsi{}}

\TitleCitation{A Precessing Jet Scenario for the Multi-Wavelength Long-Term Modulation of \lsi{}}

\Author{Fr\'ed\'eric~Jaron
$^{1,2}$\orcidA{}}

\AuthorNames{Fr\'ed\'eric~Jaron}

\AuthorCitation{Jaron, F.}

\address{%
$^{1}$ \quad Department of Geodesy and Geoinformation,
  TU~Wien, Wiedner Hauptstra{\ss}e~8,  A-1040~Vienna,  Austria; frederic.jaron@tuwien.ac.at\\
$^{2}$ \quad Max-Planck-Institut f\"ur Radioastronomie, Auf dem H\"ugel 69, D-53121 Bonn, Germany}

\abstract{The high-mass X-ray binary \lsi{} is detected across the electromagnetic spectrum from radio until the very high energy \gammaray{} regime. The emission is not only highly variable on many time scales, but is also periodic at all observed wavelengths. Periodic modulation was observed on different time-scales, ranging from hours, over months to several years. The subject of this article is a super-orbital, long-term modulation of $\sim$4.6~years. We review the observation of this periodic modulation at multiple wavelengths and investigate systematic relationships between them. In particular,  radio observations reveal that the long-term modulation is a very stable feature of the source. Observations at other wavelengths result in a phase-shift of the modulation pattern that is a systematic function of energy. The stability of this period favors a scenario in which the long-term modulation is the result of a precessing jet giving rise to periodic changes in the Doppler factor, beating with the orbital modulation of the accretion rate. We explain the phase-shifts across energy bands in a scenario with shorter wavelengths originating closer to the base of the presessing jet. A significant deviation of the TeV emission from this trend possibly requires a different explanation related to magnetic reconnection events.
}


\begin{document}

\section{Introduction} \label{sec:intro}


The high-mass X-ray binary \lsi{} is composed of a B0 Ve star \citep{Casares2005} and a black hole candidate \citep{Massi2017}.
It was discovered in 1978 during a search for variable radio sources \citep{Gregory1978}. In~the course of its observational history, spanning more than four decades now, this source has proven variable and periodic on different time scales not only in radio but all across the electromagnetic spectrum into the very high energy \gammaray{} regime. Variability time scales which have been observed in this source range from hours \citep{Peracaula1997, Jaron2017, Noesel2018, Sharma2021}, over~approximately one month (\citep[][]{Gregory2002, Jaron2018} and references therein), up~to several years (\citep[][]{Gregory2002, Massi2016, Jaron2018} and references therein). The~variability of hour time scales was observed in the radio and X-ray emission of the source and is superimposed on large outbursts which occur once per orbital cycle. This short-term (intra-day) variability is sometimes characterized by periodic oscillations, but~these are only of quasi-periodic nature \citep{Peracaula1997, Jaron2017, Noesel2018}. The~occurrence of the larger outbursts themselves, in~contrast, is periodic and precisely predictable~\citep{Jaron2013}, clearly related to the orbital period $P_1 = 26.4960 \pm 0.0028$~d of the binary system~\citep{Gregory2002}.
However, the emission from \lsi{} is periodic on an even longer time scale: both amplitude and orbital phase occurrence of the radio outbursts are modulated with a period of $P_{\rm long} = 1667 \pm 8$~d~\citep{Gregory2002}, and~there is observational evidence that the same long-term modulation is active at other wavelengths too, in~particular at X-rays~\citep{Li2012}, GeV \citep{Ackermann2013}, and~TeV \citep{Ahnen2016}. This long-term modulation 
 (\ltm{} hereafter) and its behavior at multiple wavelengths of the electromagnetic spectrum is the subject of this~article.


Since the discovery of the source\ \lsi{} the nature of the compact object in this system and the physical processes behind its non-termal emission had been debated. The~compact object has been suggested to be either a neutron star or a black hole. And~also for the physical processes two scenarios were discussed: pulsar binary and microquasar. In~the first scenario the compact object has to be a neutron star and the physical processes are powered by the spin-down of a millisecond pulsar, the~non-thermal emission being produced in the collision between the winds of that putative pulsar and the Be star (\citep[][]{Dubus2006} and references therein). This hypothesis is part of a generalization of the processes observed during the periastron passages of the pulsar binary PSR~B1259-63 \citep{Johnston1992}, occurring once every $\sim$3.4~years~(\citep[][]{Chernyakova2020} and references therein), to~a larger class of binary systems and is based exclusively on the \gammaray{} loudness of these sources \citep{Dubus2017}. As~far as \lsi{} is concernced, recent observations of the system's X-ray properties clearly show that this source fits into the class of accreting black hole binaries (\citep[][]{Massi2017} and references therein). Furthermore, its radio-loudness and the structure of its periodically occurring radio outbursts \citep{Zimmermann2015} as well as the transition from optically thick to optically thin of the radio spectral index during these outbursts \citep{Massi2009} support the presence of a jet in \lsi{} and complete the microquasar scenario. The~afore mentioned quasi-periodic intra-day variabliy at radio and X-rays has also been explained in this scenario \citep{Jaron2018, Noesel2018}. Microquasars are stellar binary systems in which a compact object accretes matter from a companion star, an~accretion disk is formed, and~a jet is launched, very similar to the processes occurring in the nuclei of radio-loud active galaxies, but~time-scales of the intrinsic physical processes are much shorter because they scale with the mass of the accretor \citep{Mirabel1998}. In~2020, \citet{Massi2020} reported an observational campaign on \lsi{} at multiple wavelengths and show how the evolution of the emission traces accretion and ejection during one orbital cycle of the binary~system.


The presence of a jet in \lsi{} is not only indirectly proven by the properties of its radio emission, but~there is also direct evidence provided by radio interferometric observations. \citet{Massi2012} analyzed a sequence of VLBA observations, clearly showing an elongated structure which changes its morphology from one-sided to two-sided and which also changes its position angle as time evolves. This not only shows that there is a visible radio jet in \lsi{} but is also an indication that this jet precesses. These authors determine a precession period of 27--28\,d, which is not equal to but larger than the orbital period of the binary system. The~value of this precession period has since been detected by timing analysis of radio \citep{Massi2013}, X-ray \citep{DAi2016}, and~GeV \citep{Jaron2014, Jaron2018} data. The~most precise value of this period comes again from VLBI observations: In 2018, \citet{Wu2018} revisited \lsi{} with VLBI astrometry and determined a value of~$P_2 = 26.926 \pm 0.005$~d.


\citet{Massi2013} in 2013 were the first authors to point out that the interference between the orbital period $P_1$ and the precession period $P_2$ results in a beating with a period identical to the period of the \ltm{}. In~this scenario, the~accretion rate onto the compact object is periodically modulated with the orbital period because of the eccentricity of the orbit \mbox{\citep{Taylor1992, Marti1995, Bosch-Ramon2006, Romero2007}}. The~precession of the relativistic jet gives rise to periodic changes in the amplification of its intrinsic emission as a result of variable Doppler boosting because of periodic changes of the angle between the jet bulk velocity and the line of sight. This scenario has been quantitatively confirmed by a physical model of a self-absorbed jet which precesses and is periodically refilled with relativistic electrons producing synchrotron emission \citep{Massi2014}. These authors show that this model reproduces the flux densities and the spectral index, both as a function of time, of~almost four decades of observational radio data. By~extending this model to include inverse Compton scattering of both internal synchrotron photons (synchrotron self-Compton) as well as external ultra-violet photons from the companion star (external inverse Compton), \citet{Jaron2016} successfully reproduce several years of simultaneous radio and GeV data resulting from long-term monitoring by the Owens Valley Radio Observatory (OVRO) at 15~GHz and the \textit{Fermi} Large Area Telescope (\fermilat{}) in the energy range 0.1--3~GeV, respectively. Analyzing continued observational data from these two ongoing monitoring programs, \citet{Jaron2018} not only showed that the presence of the two periodicities $P_1$ and $P_2$ is still evident at both wavelengths, but~furthermore detected a phase-offset in the \ltm{} patterns between the radio and GeV emission. Offsets between the phases of the \ltm{} pattern had previously been reported for other wavelengths too, as~explained in the following Section~\ref{sec:longtermobs}, but~in 2018 \citet{Jaron2018} were the first to propose a physical scenario to explain such an offset: for the radio and GeV emission, the~\ltm{} phase-offset fits into the scenario of a precessing jet in which the higher energy emission (GeV) is emitted in a jet region upstream from the position of the lower energy (radio) emission. The~reason for this interpretation is that it is the precession profile of the emission at these two wavelengths that is shifted in phase with respect to each other by an amount which agrees with the shift in the interference (i.e., \ltm{}) pattern. The~locations of these two emission regions fit into a cooling model, as~e.g.,~used by \citet{Lisakov2017} to explain the regions of the emission at these two wavelength in the jet of the active galactic nucleus~3C~273. However,~in addition, opacity effects can play a role similar to the core-shift effect at radio frequencies \citep{Lobanov1998, Pushkarev2012}.


The aim of this paper is to investigate how \ltm{} phase-offsets at other wavelengths of the emission of \lsi{} could fit into a similar scenario. For~this purpose, we are going to review how long-term observations of this source resulted in the detection of the \ltm{} at multiple wavelengths and investigate systematic relationships between these observations. The~article is structured as follows. In~Section~\ref{sec:longtermobs} we review the observational evidence for a \ltm{} being active in the source~\lsi{} at multiple wavelengths, namely at radio, optical, X-rays, GeV, and~TeV. Systematic phase-offsets between the \ltm{} pattern at different wavelengths are quantitatively investigated in Section~\ref{sec:longtermsys}. We discuss our results in the context of a physical scenario of periodic accretion and ejection in combination with jet precession in Section~\ref{sec:discussion}. We give our conclusions in Section~\ref{sec:conclusion}.

\section{Observations of the \ltm{} at Multiple~Wavelengths} \label{sec:longtermobs}

Since its first radio detection in 1978 \citep{Gregory1978}, \lsi{} has been subject of monitoring programs at different wavelengths across the electromagnetic spectrum, from~radio until the very high energy \gammaray{} regime. At~all of these wavelengths a periodic modulation of approximately 4.6~years has been detected. This section provides a review of the observational evidence for the \ltm{} at each~wavelength.

{{The following two definitions are conventionally used.}} 
 The orbital phase as a function of time $t$ is defined as
\begin{equation} \label{eq:phi}
  \Phi = \frac{t - T_0}{P_1} - \mathrm{int}\left(\frac{t - T_0}{P_1}\right),
\end{equation}
where $P_1 = 26.4960 \pm 0.0028$~d is the orbital period of the system and $T_0$ = 43,366.275~d is the MJD of the first radio detection of the source \citep{Gregory2002}. Periastron passage occurs at orbital phase $\Phi = 0.23$ \citep{Casares2005}. The~phase of the \ltm{} is defined in an analogous way, i.e.,
\begin{equation} \label{eq:theta}
  \Theta = \frac{t - T_0}{P_{\rm long}} - \mathrm{int}\left(\frac{t - T_0}{P_{\rm long}}\right),
\end{equation}
where $P_{\rm long} = 1667 \pm 8$~d is the period of the \ltm{} \citep{Gregory2002}. In~both cases $\mathrm{int}(x)$ takes the integer part of $x$. If~this term is omitted, then the phase has the meaning of orbital or long-term cycles {{elapsed}} since $T_0$, which is often useful for~calculations.

\subsection{Radio}

The \ltm{} of \lsi{} was first discovered in its radio emission. {{Already in 1987}, {\citet{Paredes1987}} {suspected in his PhD thesis that there could be a four year \ltm{}. Then i}}n 1989, based on the analysis of ten years of observational radio data in the centimeter regime, \citet{Gregory1989} reported that periodic radio outbursts, which had been known to occur once per orbit \citep{Taylor1982}, are strongly modulated in amplitude on a super-orbital, long-term time-scale. These authors {{confirm}} an approximate value of four years for the period of this \ltm{}, and~they already discussed the possibility that this long-term amplitude modulation could be the result of variable Doppler boosting caused by a similar long-term precessing jet. However, they discarded this option in favour of ``variable accretion due to quasi-cyclic Be star envelope variations'' (their Sect.~V\,b). The~reason they disfavor the precessing jet scenario is an asymmetry of the \ltm{} pattern that they observe. Such an asymmetry is indeed unexpected in a scenario which assumes the precession period of the jet to be identical to the $\sim$4~years period of the \ltm{}. However, as~we will see later on in this artice, a~precessing jet model with a precessing period close to the orbital one reproduces the observed flux densities. The~presence of the \ltm{} was then confirmed {again} in 1995 by \citet{Marti1995}, who added the observation that the orbital phase of the periodic outbursts is also affected by the \ltm{}. From~1994 until 2000, \lsi{} was part of a monitoring program of the Green Bank Interferometer (GBI) by the National Radio Astronomical Observatory (NRAO). By~observing the source several times per day, simultaneously at 2.25 and 8.3~GHz, a~database was created which remains to be highly valuable for timing analysis and the investigation of changes in the radio spectral index. In~2002, \citet{Gregory2002} applied Bayesian analysis, based on hypothesis testing, to~the GBI data and determined a refined value of $P_{\rm long} = 1667 \pm 8$~d (4.6~years) for the period of the \ltm{}\endnote{In his original paper \citet{Gregory2002} uses the symbol $P_2$ for the period of the \ltm{}. In~the present article we use the symbol $P_{\rm long}$ instead in order to avoid confusion with the precession period $P_2$ the jet in \lsi{} (\citep[][]{Wu2018} and references therein).}. In~2016, \citet{Massi2016} combined the entirety of the radio observations of \lsi{} that were available at that time. Timing analysis of this archive of 36.8~years of observational data not only confirmed the value of the long-term period, determined by \citet{Gregory2002} before, but~most importantly proved that the value of the long-term period had remained stable since the first radio detection of the source, which is eight full cycles of the \ltm{}. \citet{Massi2016} point out that variations in the circumstellar disks of Be stars, as~had been proposed by \citet{Gregory1989} but also other authors, are never observed to remain stable over such a long period of time and refer to the review paper by \citet{Rivinius2013}. {That the \ltm{} is still active in the radio emission from \lsi{} at 15~GHz (OVRO) has been shown by \citet{Jaron2018}.}

\subsection{Optical} \label{sec:opt}

First \citet{Zamanov1999} in 1999 and then \citet{Zamanov2000} in 2000 reported a \ltm{} of EW(H$\alpha$), i.e.,~the equivalent width of the H$\alpha$ emission line. \mbox{\citet{Zamanov2000}} also report a long-term phase-shift of about 0.25 with respect to the radio emission\endnote{There is an inconsistency within the numbers reported in \citet{Zamanov2000}. Namely, the~values of the phases of the fitted cosine functions in their Table~1 do not correspond to this reported \ltm{} phase-shift of 0.25. During~the review process for this article it was clarified that the value for the phase of EW(H$\alpha$) is in fact $\phi_0 = 0.37 \pm 0.01$ (and not $0.60 \pm 0.01$). Hence, the~\ltm{} phase-shift quoted in the abstract of \citet{Zamanov2000} remains correct.}. In~2015 \citet{Paredes-Fortuny2015} reported on a phase-shift of the orbital phase occurrence of peaks of HW(H$\alpha$), a~phenomenon that is known to be part of the \ltm{} at radio and GeV wavelengths (\citep[][]{Jaron2014} and references therein). However, the~conclusions by \mbox{\citet{Paredes-Fortuny2015}} are based on only 1.5~years of observational data, i.e.,~only about one third of the 4.6~years \ltm{}.

A direct comparison of the 0.25 \ltm{} phase-offset reported by \citet{Zamanov2000} to the other phase-shifts reported in our present investigation is complicated by the fact that they used a different value of 1584~d for the period of the \ltm{}. Therefore, before~these data could be used for our investigation they would have to be re-analyzed taking into account current knowledge about the timing characteristics of the emission from \lsi{}, in~particular the value \citep{Gregory2002} and the stability \citep{Massi2016, Jaron2018} of the \ltm{} period. However,~there is also a more fundamental reason that leads us to the decision not to include any EW(H$\alpha$) data: it would require a thorough investigation to disentangle the contributions from different emission regions within the stellar binary system to EW(H$\alpha$). Such an investigation is beyond the scope of this article, but~a discussion of this issue in connection with optical emission regions will be given in Section~\ref{sec:optdisc}.

\subsection{X-rays}

X-ray emissions from \lsi{} have been detected since 1981 \citep{Bignami1981}, but~the first authors to report the \ltm{} in this energy range were \citet{Li2012} in 2012. They analyzed four years of \textit{RXTE}-PCA data of the energy range 3-30~keV. In~the same publication they report a shift between the \ltm{} at X-rays compared to the radio emission. They quantify this shift to $281.8 \pm 44.6$~d, which translates to $0.17 \pm 0.03$ in terms of long-term phase-offset with respect to the radio emission, meaning that the X-ray modulation lags the radio. The~left panel of their Figure~1, showing the raw peak count rates put into long-term phase bins, reveals that the envelope of the \ltm{} pattern in the X-rays has the same asymmetry as has also been observed in the radio. It has the characteristic shape of a steep rise and a slower decline. The~phase-offset cited here was obtained by the authors by fitting a sine wave to the ''modulated flux fraction'' (MFF). The~effect of this MFF is that indeed the long-term modulation of it looks more sinusoidal than the original X-ray count peaks. In~reality the original count peaks look much more similar to the radio data. Indication that the mechanism behind the \ltm{} in the X-ray emission is the beating between orbit and precession comes from the fact that \citet{DAi2016} detected the corresponding spectral features of orbital ($P_1$) and precession ($P_2$) periods in the light curves of Swift/BAT survey~data.

In 2014, \citet{Li2014} analyzed \textit{INTEGRAL} X-ray data from a slightly higher energy interval (18--60\,keV). They detect the \ltm{} also in this energy range and report that the \ltm{} pattern has a peak at $\Theta \approx 0.2$. Since the \ltm{} at radio has its maximum at \mbox{$\Theta = 0.94 \pm 0.02$}~\citep{Jaron2018}, this corresponds to a phase-shift of $\sim$0.26. However, because~this is only a rough estimate without reported uncertainty and because the energy range overlaps with the previously cited \citet{Li2012} we do not include this data point in our quantitative analysis later on in this~article.

\subsection{GeV}

Since the source had been detected in the radio \citep{Gregory1978}, the~possibility of \gammaray{} emission from \lsi{} was discussed as early as in 1979 \citep{Gregory1979}. A~firm association with a GeV source, previously detected by \citet{Hermsen1977} in 1977, could only be established in 2009, when \citet{Abdo2009} detected the orbital period in the emission observed by the \fermilat{}. The~\fermilat{} has also been the first mission to provide continuous monitoring in the GeV regime, and~in this way, it enabled the investigation of the long-term behavior of the source in this energy~range.

Other than in the radio, where the \ltm{} is already evident by an eye-inspection of the raw observational data, detecting the same modulation in the GeV required a closer look at the data. By~analyzing several years of \fermilat{} data but selecting only certain orbital phase ranges, in~2013 \citet{Ackermann2013} detected for the first time a significant long-term amplitude modulation which, however, only affects apastron orbital phases (\mbox{$\Phi$ = 0.5--1.0}). \citet{Jaron2014} found out that the GeV emission has indeed two distinct peaks along the orbit. The~first peak occurs at periastron and remains stable in orbital phase. The~second peak occurs at a later orbital phase and is subject to the same phase-shift as observed for the radio outbursts. In~combination, the~results obtained by \mbox{\citet{Ackermann2013}} and \citet{Jaron2014} completed the picture that the \ltm{} has the same period and the same effects, in~terms of amplitude modulation and orbital phase-shift of the outbursts, as~is the case for the radio emission. In~addition to that, there is a GeV outburst at periastron, which remains stable in amplitude and orbital phase, while the radio emission is lacking a distinct periastron~outburst.

\textls[-25]{The \ltm{} remains active in {both the radio and} the GeV regime, as~shown by \mbox{\citet{Jaron2018}}. Furthermore, these authors detect a phase-shift between the \ltm{} at radio and GeV, which they quantify to $0.26 \pm 0.03$\endnote{In their original publication, \citet{Jaron2018} determine the phase-offset between the \ltm{} pattern at radio relative to the GeV emission. This is why their value has the opposite sign. In~this present work, we express all \ltm{} phase-offsets with respect to the radio emission. Since all higher energy emissions \emph{lag} the radio in term of the \ltm{} pattern, this results in positive signs for the phase-offsets reported here.}. As an explanation for this phase-offset they suggest a scenario in with the GeV emission is produced in a precessing jet upstream from the radio emission. We will come back to this scenario and extend it to the multi-wavelength picture further down in this~article.}

\subsection{TeV}


The source \lsi{} is also detected in the very high energy (VHE) \gammaray{} regime. Here we first summarize the history of observations of the source with Cherenkov telescopes. As~we will see, the~\ltm{} has been detected in the TeV emission, but~a quantitative determination of its phase-offset and uncertainties of the fit parameters has so far been missing from the literature. In~order to obtain these values for our investigation here, we repeat a previously published analysis in the second~part.

\subsubsection{Observational~History}

A VHE counterpart to \lsi{} was first detected by \citet{Albert2006}. The~source has since been included in the monitoring by MAGIC and VERITAS \citep[see references in][]{Ahnen2016}. The~\ltm{} of \lsi{} was detected in the VHE \gammaray{} regime by \citet{Ahnen2016} in 2016 by combining observations from both of these VHE observatories. Selecting only peak flux densitities from the orbital phase interval $\Phi$ = 0.50--0.75, they showed that these data are best fit with a sine function. They report a period of this sine of $1610 \pm 58$~d, which agrees well within uncertainties with the value of $1667 \pm 8$~d determined from radio observations~\citep{Gregory2002}. However, \citet{Ahnen2016} do not mention the values of the other fit~parameters.

\subsubsection{Determination of the \ltm{} Phase-Offset} \label{sec:dettev}

\begin{figure}[H]
  \includegraphics{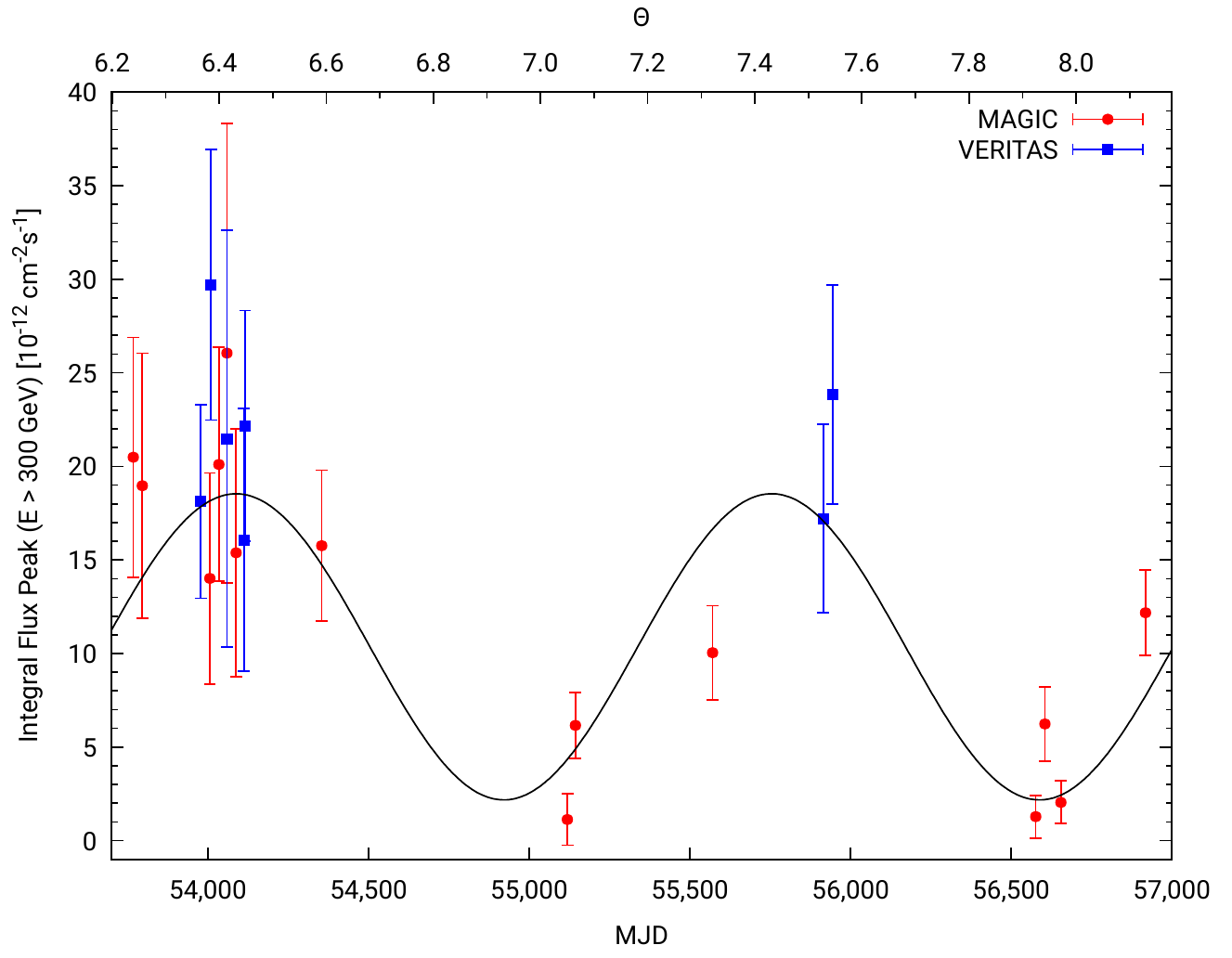}
  \caption{
    VHE \gammaray{} data resulting from MAGIC (reda points) and VERITAS (blue squares) observations. The~data for this plot were taken from \citet{Ahnen2016} and correspond to the peak fluxes of each orbital cycle (for details see their Section~3.2), plotted against time (MJD, lower axis) and long$-$term phase $\Theta$ (upper axis). The~solid black curve is the result of fitting these data with a sine function, as~described here in Section~\ref{sec:dettev}.
  }
 \label{fig:TeV}
\end{figure}

To determine the value of the long-term phase-offset in the TeV regime, we repeat here the analysis of \citet{Ahnen2016}. We take the data that they present in their Figure~2, plotted in Figure~\ref{fig:TeV} here. We fit these data with a sine function of the form
\begin{equation} \label{eq:sine}
  f(\Theta) = a\sin 2\pi(\Theta - \Theta_0) + b,
\end{equation}
which is a function of the long-term phase $\Theta$, with~amplitude~$a$, offset~$b$, and~long-term phase-origin~$\Theta_0$. The~long-term phase $\Theta$ is defined in Equation~(\ref{eq:theta}). We keep the value of the long-term period~$P_{\rm long} = 1667$\,d fixed to the value from the radio. This approach is justified by the fact that the value determind by \citet{Ahnen2016} for the TeV agrees with the value for the radio. In~addition to that, this enables the comparison between a phase-offset between TeV and radio, i.e.,~the subject of this~article.

\begin{figure}[H]
  \includegraphics{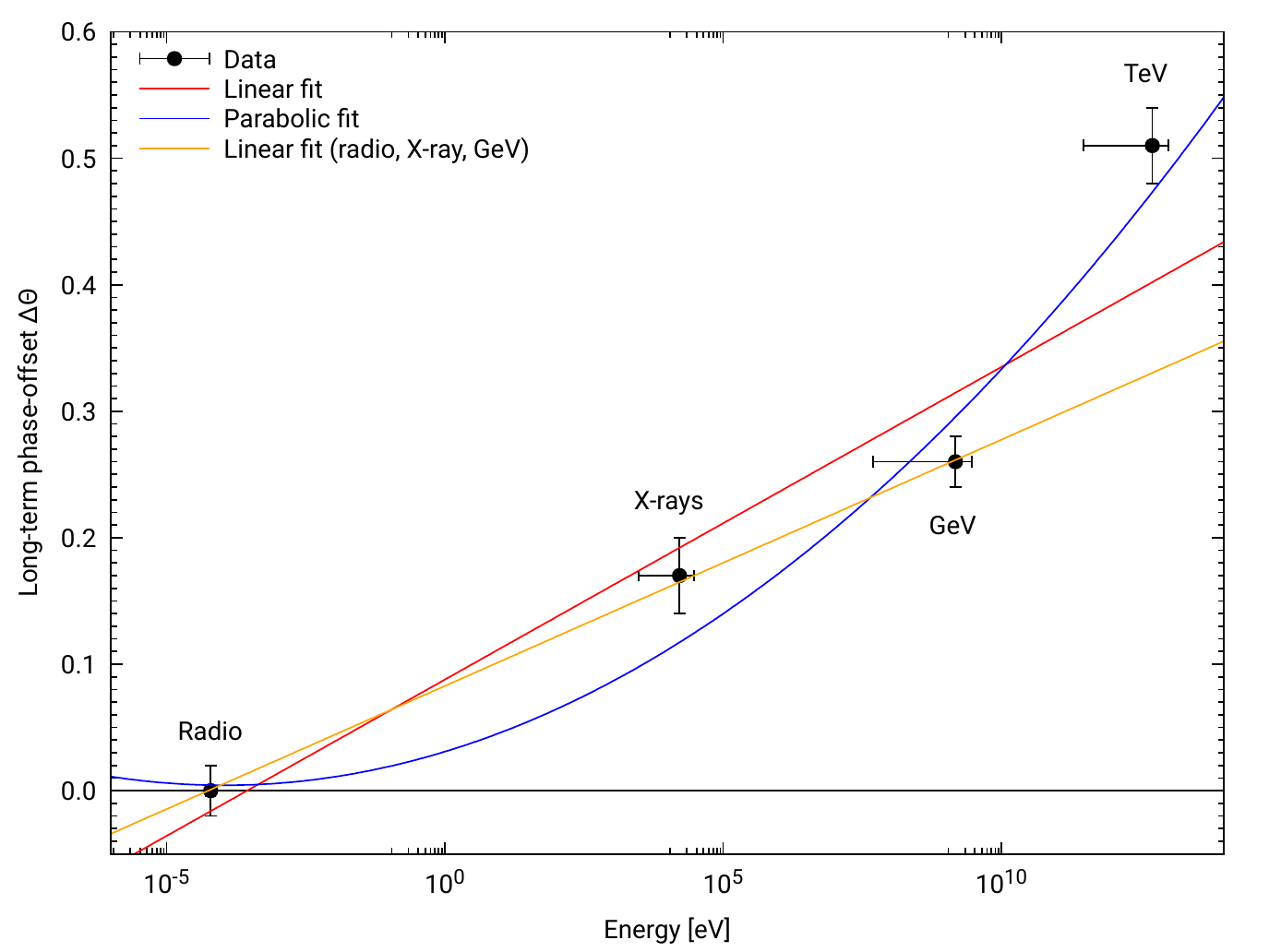}
  \caption{
    Phase$-$offset of the long-term modulation of \lsi{} as a function of the energy of the observed emission. Vertical error bars represent the $1\sigma$ uncertainties of the determined phase-offset. Horizontal error bars show the energy range of the data taken to determine that value. We define the phase of the long-term modulation as $\Delta \Theta = \Theta - \Theta_{\rm radio}$. The~solid curves are phenomenological fits to the data, as~indicated by the line colors and explained in Section~\ref{sec:longtermsys}.
  }
  \label{fig:phase}
\end{figure}

We confirm that the data are well fitted by the sine function of the form of \mbox{Equation~(\ref{eq:sine})}. The~reduced $\chi^2$ from our analysis is $1.6$, and~the values of the fit parameters are listed in Table~\ref{tab:sinefit}. We obtain a value of $\Theta_0 = 0.20 \pm 0.03$ for the phase-origin of the \ltm{} in \lsi{} at TeV energies. {Relative to the phase-origin of the \ltm{} at radio $\Theta_0 = 0.69$ determined by \citet{Jaron2018} (called $\phi_0$ in the last row of their Table~2), t}his translates to a phase-offset of $0.51 \pm 0.03$ between the \ltm{} at TeV and radio. At~this point there is the question of ambiguity of the phase-offset. The~same offset could also have been interpreted as $-0.49$, i.e.,~the TeV being earlier than the radio by that amount. However, as~evident from Figure~\ref{fig:phase}, a~positive TeV-radio offset leads to the best alignment with the offsets at the other wavelengths. {As we will see in Section~\ref{sec:cool}, this value of $0.51$ for the TeV phase-offset also fits into a physical scenario in which these \ltm{} phase-offsets correspond to energy-dependent emission regions in a precessing jet. However, as~we will see in Section~\ref{sec:mr}, the~TeV photons may be produced by a different mechanism, and~the ambiguity resolution of this phase-offset value would then not be so obvious any more.}

\begin{specialtable}[H]
  \caption{
    Values of the parameters resulting from fitting the TeV data of Figure~\ref{fig:TeV} with a sine function of the form given in Equation~(\ref{eq:sine}).
  }
  \label{tab:sinefit}
  \setlength{\tabcolsep}{19.4mm} 
  \begin{tabular}{lc}
    \toprule
    \textbf{Parameter} & \textbf{Value}\\
    \midrule
    $a$                & $(8.18 \pm 1.11)~10^{-12}\mathrm{cm}^{-2}\mathrm{s}^{-1}$\\
    $b$                & $(10.36 \pm 1.11)~10^{-12}\mathrm{cm}^{-2}\mathrm{s}^{-1}$\\
    $\Theta_0$         & $0.20 \pm 0.03$\\
    \bottomrule
  \end{tabular}  
\end{specialtable}
\unskip

This concludes our review of observations of the \ltm{} in \lsi{} at multiple wavelengths. In~the next section, we are going to put the relative phase-offsets in context and investigate systematics between~them.

\section{Phase-Offsets across~Wavelengths} \label{sec:longtermsys}

All the known values of the \ltm{} phase-offsets at different wavelengths of the emission from \lsi{} are listed in Table~\ref{tab:phase}. The~energy range reported there is the one of the emission used to determine the phase-offset at that wavelength. The~\ltm{} pattern of the higher energy emission \emph{lags} the pattern of the radio emission, the~offset of which we define as zero. Figure~\ref{fig:phase} presents a plot of the values given in Table~\ref{tab:phase}. Plotted is long-term phase-offset against energy of the observed emission, the~energy axis appearing in a logarithmic scale. It is evident from this plot that the phase-offset monotonically increases as a function of energy, meaning that the lag of the \ltm{} pattern, with~respect to radio, increases with energy. At~this point it has to be mentioned that there is an ambiguity affecting all of these offsets. For~example, the~value for the X-rays could also be interpreted as $-0.83$, i.e.,~leading the radio emission by that amount, and~analogous for the other wavelengths. However, resolving these ambiguities to the values reported here results in the smallest \ltm{} phase-offsets in between adjacent wavelengths, and~it also agrees with the conclusions given in previous publications \citep{Li2012, Jaron2018}, with~\citet{Jaron2018} even providing a physical~explanation.

\begin{specialtable}[H]
  \caption{
    Phase-offsets of the long-term modulation, as~observed at different wavelengths of the emission from \lsi{}. All phase-offset values are given relative to the long-term phase of the radio emission. The~higher energy emission lags the radio in terms of \ltm{}.
  }
  \label{tab:phase}
  \setlength{\tabcolsep}{3.1mm} 
  \begin{tabular}{llll}
    \toprule
    \textbf{Name} & \textbf{Energy Range [eV]} & \textbf{Phase-Offset} & \textbf{Reference}\\
    \midrule
    Radio  & (5.0--7.4) $\times\,10^{-5}$ & $0 \pm 0.02$ & \citet{Jaron2018}\\
    X-rays (soft) & (3--30) $\times\,10^{3}$ & $0.17 \pm 0.03$ & \citet{Li2012}\\
    X-rays (hard) & (18--60) $\times\,10^{3}$ & $\sim$0.2 & \citet{Li2014} (not used here)\\
    GeV    & (0.1--3.0) $\times\,10^{9}$  & $0.26 \pm 0.02$ & \citet{Jaron2018}\\
    TeV    & (0.3--10) $\times\,10^{12}$  & $0.51 \pm 0.03$ & This work, based on \citep{Ahnen2016}\\
    \bottomrule
  \end{tabular}
\end{specialtable}

To quantify the relationship between \ltm{} phase-offset and the logarithm of the energy, we fit the data presented in Figure~\ref{fig:phase} with different functions. The~first function is a straight line of the form
\begin{equation} \label{eq:lin}
  f_1(E) = a_1\cdot\log_{10}E + b_1,
\end{equation}
where the energy $E$ is given in eV. This fit results in the parameters $a_1 = 0.02 \pm 0.01$, $b_1 = 0.09 \pm 0.05$, and~a reduced $\chi^2/\mathrm{d.o.f.} = 21.6/(4 - 2) = 10.8$. This linear fit is plotted as the red line in Figure~\ref{fig:phase}.  This shows that the slope $a$ is clearly positive, quantitatively confirming the increasing offset with increasing energy. However,~the large $\chi^2$ means that the relationship is certainly not that~simple. 

To account for a possible curvature in the trend, the~second function is a parabola of the form
\begin{equation} \label{eq:sqr}
  f_2(E) = a_2\cdot\left(\log_{10}E\right)^2 + b_2\cdot\log_{10}E + c_2,
\end{equation}
\textls[-20]{and the parameters are estimated to $a_2 = 0.002 \pm 0.001$, $b_2 = 0.013 \pm 0.010$, \mbox{$c_2 = 0.031 \pm 0.058$}, and~a reduced $\chi^2/\mathrm{d.o.f.} = 7.8/(4 - 3) = 7.8$. The~blue line in Figure~\ref{fig:phase} shows that this fit represents the data actually quite well despite the rather high $\chi^2$. The~reason for this is the small number of available data~points.}

Finally, we note that the data points for the radio, X-ray, and~GeV emission seem to lie on a straight line, while the TeV point is apparently offset from this otherwise linear trend. In~order to quantify this impression, we restrict the linear fit of \mbox{Equation~(\ref{eq:lin})} to the first three data points. This results in fit parameters $a_1 = 0.0195 \pm 0.0004$, $b_1 = 0.0828 \pm  0.0027$, and~a reduced $\chi^2/\mathrm{d.o.f.} = 0.04/(3 - 2) = 0.04$. The~corresponding orange line in \mbox{Figure~\ref{fig:phase}} shows that the three first data points happen to perfectly lie on this line, while the TeV point is clearly offset. The~value of the fitted line at the position of the TeV is $f_1(\unit[5.15]{TeV}) = 0.33$, and~the TeV measurement data point is with $0.51 \pm 0.03$ offset from that value at the $6\sigma$ level. {If the phase-ambiguity is resolved to a value of $-0.49$ instead, then this deviation has a significance of even $27\sigma$.}

Concerning the function fitting described above it is important to point out that the small number of data points greatly restricts the possibilities here. In~the time of writing, we only have four data points to work with. In~order to draw more solid conclusions from this, Table~\ref{tab:phase} and the corresponding plot in Figure~\ref{fig:phase} have to be populated with more data points. The~quantitative findings of systematic \ltm{} phase-offsets as a function of energy will be discussed in a physical scenario in the next~section.

\section{Discussion} \label{sec:discussion}

The \ltm{} in the emission from \lsi{} has been detected at multiple wavelengths. As~evident from the observational facts presented in Section~\ref{sec:longtermobs}, the~phase of the modulation pattern is not the same at all wavelengths, but~is significantly offset from each other. Putting these phase-offsets into context, as~we did in Section~\ref{sec:longtermsys}, reveals that this phase-offset is monotonically increasing as the energy of the observed emission increases, as~evident from Figure~\ref{fig:phase} and the functional fits to these data. A~physical scenario that explains the observed long-term phase-offset between the emissions at radio and GeV wavelengths has already been descibed by \citet{Jaron2018}. In~the following we first show that this scenario can be straightforwardly extended to also include the phase-offsets in the X-ray and TeV energy regimes. However, the~divergence of the TeV data point in Figure~\ref{fig:phase} from the otherwise very simple straight line through the data points at the other three wavelengths deserves an extra comment, given in the second part of this section. {Finally, in~the third part we will further comment on the exclusion of optical data and describe how their inclusion could help to further discriminate between emission models for this source.}


\subsection{Plasma Cooling and Opaticity in a Precessing~Jet} \label{sec:cool}

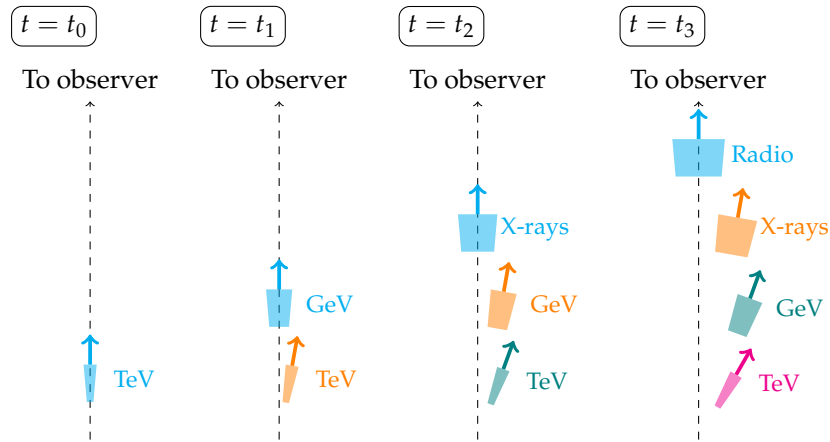
\begin{figure}[H]
\begin{tikzpicture}
  \fill (-0.481187649392582,5.5) node[draw, rounded corners] {$t = t_{0}$};
  \draw[->,dashed] (0,0) -- (0,4.5);
  \draw (0,4.8) node {To observer};
  \fill [opacity=0.5,cyan] (-0.043744331762962,0.5) -- (-0.0873222031367911,0.998097349045873) -- (0.0873222031367911,0.998097349045873) -- (0.043744331762962,0.5) -- cycle;
  \draw[cyan, ->, line width=0.5mm] (0,0.998097349045873) -- (0,1.39809734904587);
  \draw (0.587322203136791,0.8) node {\small \textcolor{cyan}{TeV}};
\end{tikzpicture}
  \quad
\begin{tikzpicture}
  \fill (-0.481187649392582,5.5) node[draw, rounded corners] {$t = t_{1}$};
  \draw[->,dashed] (0,0) -- (0,4.5);
  \draw (0,4.8) node {To observer};
  \fill [opacity=0.5,orange] (0.043744331762962,0.5) -- (0.0873222031367911,0.998097349045873) -- (0.259313368455229,0.967770666156742) -- (0.129903845903968,0.484807753012208) -- cycle;
  \draw[orange, ->, line width=0.5mm] (0.17331778579601,0.982934007601308) -- (0.242777056862782,1.37685710880619);
  \draw (0.759313368455229,0.8) node {\small \textcolor{orange}{TeV}};
  \fill [opacity=0.5,cyan] (-0.131232995288886,1.5) -- (-0.174810866662715,1.99809734904587) -- (0.174810866662715,1.99809734904587) -- (0.131232995288886,1.5) -- cycle;
  \draw[cyan, ->, line width=0.5mm] (0,1.99809734904587) -- (0,2.39809734904587);
  \draw (0.674810866662715,1.8) node {\small \textcolor{cyan}{GeV}};
\end{tikzpicture}
  \quad
\begin{tikzpicture}
  \fill (-0.481187649392582,5.5) node[draw, rounded corners] {$t = t_{2}$};
  \draw[->,dashed] (0,0) -- (0,4.5);
  \draw (0,4.8) node {To observer};
  \fill [opacity=0.5,teal] (0.129903845903968,0.484807753012208) -- (0.259313368455229,0.967770666156742) -- (0.42342542829205,0.908038761292025) -- (0.2121162974217,0.4548848677737) -- cycle;
  \draw[teal, ->, line width=0.5mm] (0.341369398373639,0.937904713724384) -- (0.478177455703907,1.31378176203875);
  \draw (0.92342542829205,0.784807753012208) node {\small \textcolor{teal}{TeV}};
  \fill [opacity=0.5,orange] (0.131232995288886,1.5) -- (0.174810866662715,1.99809734904587) -- (0.519121060263165,1.93738617218116) -- (0.389711537711905,1.45442325903662) -- cycle;
  \draw[orange, ->, line width=0.5mm] (0.34696596346294,1.96774176061352) -- (0.416425234529712,2.3616648618184);
  \draw (1.01912106026317,1.8) node {\small \textcolor{orange}{GeV}};
  \fill [opacity=0.5,cyan] (-0.21872165881481,2.5) -- (-0.262299530188639,2.99809734904587) -- (0.262299530188639,2.99809734904587) -- (0.21872165881481,2.5) -- cycle;
  \draw[cyan, ->, line width=0.5mm] (0,2.99809734904587) -- (0,3.39809734904587);
  \draw (0.762299530188639,2.8) node {\small \textcolor{cyan}{X-rays}};
\end{tikzpicture}
  \quad
\begin{tikzpicture}
  \fill (-0.481187649392582,5.5) node[draw, rounded corners] {$t = t_{3}$};
  \draw[->,dashed] (0,0) -- (0,4.5);
  \draw (0,4.8) node {To observer};
  \fill [opacity=0.5,magenta] (0.2121162974217,0.4548848677737) -- (0.42342542829205,0.908038761292025) -- (0.574671920753823,0.820716558155234) -- (0.2878837025783,0.411140536010738) -- cycle;
  \draw[magenta, ->, line width=0.5mm] (0.499048674522936,0.86437765972363) -- (0.699048674522936,1.21078782123741);
  \draw (1.07467192075382,0.7548848677737) node {\small \textcolor{magenta}{TeV}};
  \fill [opacity=0.5,teal] (0.389711537711905,1.45442325903662) -- (0.519121060263165,1.93738617218116) -- (0.847658023135451,1.81780849683943) -- (0.636348892265101,1.3646546033211) -- cycle;
  \draw[teal, ->, line width=0.5mm] (0.683389541699308,1.87759733451029) -- (0.820197599029575,2.25347438282466);
  \draw (1.34765802313545,1.75442325903662) node {\small \textcolor{teal}{GeV}};
  \fill [opacity=0.5,orange] (0.21872165881481,2.5) -- (0.262299530188639,2.99809734904587) -- (0.778928752071102,2.90700167820557) -- (0.649519229519842,2.42403876506104) -- cycle;
  \draw[orange, ->, line width=0.5mm] (0.520614141129871,2.95254951362572) -- (0.590073412196643,3.34647261483061);
  \draw (1.2789287520711,2.8) node {\small \textcolor{orange}{X-rays}};
  \fill [opacity=0.5,cyan] (-0.306210322340734,3.5) -- (-0.349788193714563,3.99809734904587) -- (0.349788193714563,3.99809734904587) -- (0.306210322340734,3.5) -- cycle;
  \draw[cyan, ->, line width=0.5mm] (0,3.99809734904587) -- (0,4.39809734904587);
  \draw (0.849788193714563,3.8) node {\small \textcolor{cyan}{Radio}};
\end{tikzpicture}

  \caption{
    Sketch of the precessing jet scenario in the multi-wavelengths context. The~figures appear in chronological order from left to right, each representing an epoch during which a new ejection of plasma occurs. Each ejection is marked with a different color and propagates, as~indicated by the colored arrows, in~a different direction with respect to the line of sight as time evolves. Expansion of the plasma is represented by the ejections becoming broader, following a conical shape, as~they propagate. The~energy regime of the emission from each ejection is labelled right to it in the same color.
  }
  \label{fig:sketch}
\end{figure}

Figure~\ref{fig:sketch} presents a sketch visualizing how the phase-offsets between the multiple wavelengths fit into a precessing jet scenario. This figure is basically an extension of Figure~7 in \citet{Jaron2018}, which showed an analogous sketch, but~restricted to the radio and GeV emissions. Figure~\ref{fig:sketch} shown here consists of four sub-figures which represent four epochs, ordered chronologically from left to right and labelled $t_0$ to $t_3$. The~interval $t_0-t_3$ is short compared to the orbital period $P_1${, with~a separation of a few days between adjacent time steps. As~a particular example, the~results in \citet{Jaron2018} would translate to a difference of about five days between $t_1$ and $t_3$}. The~line of sight is drawn as a dashed arrow which points into the direction of the observer. Different ejections of plasma into a conically expanding jet are marked with different colors as they appear. The~ejections propagate into the direction indicated by an arrow and emit radiation at an energy regime as labelled in the same color as the plasma itself. At~time~$t_0$ an ejection of plasma appears and is marked in blue. It propagates into a direction that is aligned with the line of sight\endnote{In reality it is not necessarily the case that the jet is perfectly aligned with the line of sight at any time. However,~there is certainly a moment when the jet encloses the smallest angle with respect to the line of sight. For~the sake of clarity, in~Figure~\ref{fig:sketch} this smallest angle is chosen to be zero.}. The~emission from this plasma is in the TeV regime. Because~it travels at a significant fraction of the speed of light ($\beta \approx 0.5$, \citet{Jaron2016}), the~intrinsic emission appears amplified to the observer as a result of maximal Doppler boosting. At~$t_1$ the blue ejection has propagated on a ballistic trajectory, following its initial direction and expanding in a conical manner. The~energy range of the emission of the blue plasma is now in the GeV regime, as~a result of plasma cooling. At~the same time, i.e.,~still at $t_1$, a~new ejection of plasma appears, which is now marked in orange. Because~of the precession of the jet, this ejection now propagates into a slightly rotated direction compared to the previous blue ejection and is now inclined with respect to the line of sight. The~emission of the orange plasma is in the TeV regime, but~the Doppler boosting is now decreased compared to the TeV emission of the blue plasma at time~$t_0$. In~the next step, at~time $t_2$, the~blue plasma has continued its propagation and expansion, and~now emits in the X-ray regime. The~orange plasma has done the same in its own direction, now emitting GeV. In addition,~there is also a new plasma, this time marked in green, and~which is ejected into a yet increased angle as a result of continuous jet precession. Finally, at~$t_3$, the~multi-wavelengths picture is complete with the original blue plasma having cooled down to emit in the radio. The~orange and green ones have continued the process to emit in X-rays and GeV, respectively. In addition,~a newly appearing magenta plasma emits in the TeV. All of these emissions at different wavelengths originate from plasmas moving into different directions with respect to the line of sight and hence appear differently Doppler boosted to an observer, who remains at a fixed~position.

The process described above shows that the emission at each energy regime (i.e., TeV, GeV, X-rays, and~radio) has its peak of amplification at a different moment in time in the interval $t_0$--$t_3$. Following the blue plasma, the~maximum Doppler boosting for the TeV emission occurs at $t_0$, for~GeV at $t_1$, for~X-rays at $t_2$, and~for radio at $t_3$. Looking at the temporal evolution of the direction of one wavelength alone, e.g.,~TeV, one can trace how the amplification at this energy changes with time. We refer to this variability in the Doppler boosting of the intrinsic emission as the \emph{precession profile} at a certain wavelength. The~crucial point to emphasize now is that the peak of maximal Doppler boosting occurs \emph{earlier} in the precession profile with increasing energy. This is the complete opposite as for the \ltm{} pattern, where the peak of the higher energy occurs \emph{later}. In~the following we will see how this can be understood by investigating mathematically the interference of two periodic~processes.

Mathematically, the~precession profile can be expressed, to~first order, as~a cosine function of the form
\begin{equation} \label{eq:pp}
  f_{\rm prec}(t) = \cos 2\pi\left(\frac{t - T_0}{P_2} - \phi_{\rm mp}\right),
\end{equation}
where $P_2$ is the precession period of the jet and $T_0$ is the conventional time-origin\endnote{Please note that we denote this time-origin with a capital $T_0$, not to be confused with the $t_0$ used in Figure~\ref{fig:sketch} and the explanation~above.} for \lsi{} (\citep[][]{Jaron2018} and references therein). The~phase~$\phi_{\rm mp} = \phi_{\rm mp}(E)$ of maximum (m) Doppler boosting due to precession (p) is a function of the energy~$E$ of the observed emission. This has been empirically proven for the radio and GeV emission by \mbox{\citet{Jaron2018}}, and~our scenario here is based on the assumption that this is also true for the X-ray and TeV emission. We follow the convention that the phase~$\phi_{\rm mp}$ is a number between 0 and 1, as~for the orbital~phase.


Several authors showed that for the eccentric orbit of \lsi{} two accretion peaks are predicted \citep{Taylor1992, Marti1995, Bosch-Ramon2006, Romero2007}. As~pointed out by \citet{Taylor1992}, the~\citet{Bondi1952} accretion rate $\dot{M} \propto \rho/v_{\rm rel}^3$ is proportional to the density~$\rho$ of the stellar wind at the position of the accretor, and~inversly proportional to the cube of the relative velocity~$v_{\rm rel}$ between the accretor and the wind. As~a consequence, one accretion peak is at periastron, where the density~$\rho$ is maximal. If~the eccentricity~$e$ of the orbit is sufficienty high, there is a second accretion peak at a later orbital phase, when the smaller velocity~$v_{\rm rel}$ compensates for the lower density~$\rho$. As~\citet{Taylor1992} showed (see their Figure~7), this second accretion peak develops at an eccentricity~$e = 0.4$ and is well pronounced for $e = 0.6$. The~eccentricity of the orbit of \lsi{} has been determined to be $e = 0.72 \pm 0.15$ by \mbox{\citet{Casares2005}}, so two peaks along the orbit are clearly expected. It is, however, only the emission from the second, apastron, accretion peak which is subject to the \ltm{}. \mbox{\citet{Jaron2016}} explain this observational fact with a physical model in which the ejection at periastron only reaches velocities of $\beta \approx 0.01$, which is not fast enough to give rise to appreciable Doppler boosting. For~this reason, we restrict the accretion profile here to apastron orbital phases. Mathematically this leads to an expression analogous to the one given in Equation~(\ref{eq:pp}),
\begin{equation} \label{eq:op}
  f_{\rm orb}(t) = \cos 2\pi\left(\frac{t - T_0}{P_1} - \phi_{\rm mo}\right),
\end{equation}
where $P_1$ is the well-known orbital period of the system and $T_0$ is the same time-origin as for the precession profile above. The~phase $\phi_{\rm mo}$ of the maximum~(m) of the orbital~(o) modulation of the accretion rate does \emph{not} depend on the energy, but~is a constant \mbox{$\phi_{\rm mo} = \mathrm{const}$}, extrapolating the results from \citet{Jaron2018} to the X-rays and TeV. (Following \mbox{\citet{Jaron2016}} this is $\phi_{\rm mo} = 0.58$, see their Table~1, called $\Phi_0$ for injection II there.) We point out that the actual value of $\phi_{\rm mo}$ is not important for the following explanation, so we are going to neglect this term from here on for the sake of clarity of the~formulas. 


The emission from \lsi{}, as~received by an observer on the Earth, is modulated by two mechanism. The~first mechanism is the modulation of the accretion rate, and~the second one is the amplification of the intrinsic emission of the jet resulting from periodic changes in the Dopper boosting due to precession of the relativistic jet. These two mechanisms interfere with each other, giving rise to an interference pattern that has the form of a beating because of the proximity of the two intrinsic periods $P_1$ and $P_2$. This beating is identical to the \ltm{}, as~first discovered by \citet{Massi2013}. This hypothesis has been confirmed by physical modelling of the radio \citep{Massi2014} and GeV \citep{Jaron2016} data, resulting from long-term monitoring of the source. Here we investigate the effect of an energy-dependent phase-offset of the precession profile in a way that is analogous to the presentation in the appendix of \citet{Jaron2018}. The~interference between orbit and precession can be studied by considering the sum of the two,
\begin{eqnarray}
   f_{\rm beat}(t)  =  f_{\rm orb}(t) + f_{\rm prec}(t)
   =  \cos 2\pi\left(\frac{t - T_0}{P_1}
   \right) + \cos 2\pi\left(\frac{t - T_0}{P_2} {- \phi_{\rm mp}}\right) \nonumber\\
   \propto \cos2\pi\left(\frac{t - T_0}{P_{\rm avg}}
   - \frac{\phi_{\rm mp}}{2}\right) \cos2\pi\left(\frac{t - T_0}{2P_{\rm beat}}
   {+ \frac{\phi_{\rm mp}}{2}}\right),
\end{eqnarray}
where the first cosine term oscillates at a period which is the inverse of the average of the frequencies $1/P_1$ and $1/P_2$, and~the second cosine term is slowly oscillating at the twice the beat period $P_{\rm beat} = 1/(P_1^{-1} - P_2^{-1})$. The~envelope of this interference pattern has a period of $P_{\rm beat}$ and is identical to the \ltm{} of \lsi{}, i.e.,~$P_{\rm beat} = P_{\rm long}$, as~explained in \citet{Massi2013}. The~important point here is that the precession phase-offset $\phi_{\rm mp}$ propagates into the interference pattern with the \emph{opposite} sign. This means that emission which has its maximum \emph{earlier} in the precession profile (corresponding to \emph{negative} values of $\phi_{\rm mp}$) has its maximum in the \ltm{} pattern shifted to a \emph{later} time (or phase). This is exactly what we observe for the \ltm{} phase-offsets presented in Figure~\ref{fig:phase}. Remembering that the \ltm{} pattern corresponds to the envelope of the interference pattern, the~phase shift of the \ltm{} occurs exactly by the same amount as for the precession profile but in the opposite direction, as~outlined by \citet{Jaron2018}.

In this way, we can conclude that the sequence of the emission of the energy regimes TeV, GeV, X-ray, and~radio occurs in that particular order in the precessing jet scenario. However, it is important to point out that this reflects a temporal order first of all. Because~the jet is filled with plasma at its base and material propagates in the jet direction from that point on, we can also conclude that this temporal order is identical to the spatial sequence of emission regions along the jet. However, to~draw conclusions about the exact locations of these regions is not possible in this scenario and with the data available. The~reason for this is that there is a degeneracy of the problem in terms of distance from the base and velocity of the plasma along the jet. It is well possible that the processes probed by the emission observed for this study occur in a part of the jet where acceleration of the plasma still plays an important role. This has also been pointed out by \citet{Jaron2016} to explain the slow velocity of a jet that is launched at periastron of the system, and~which is not affected by the \ltm{} because the bulk velocity does not reach relativistic~speeds.

The deviation of the TeV data point in Figure~\ref{fig:phase}, however, is remarkable. In~the scenario presented until here there are three possible reasons for this relatively large~offset:
\begin{enumerate}
\item{
  The distance between the location of the TeV and GeV emission is considerably larger than between the other wavelengths.
}
\item{
  It takes the plasma longer to travel from the TeV region to the GeV region than in between the other regions because the velocity is smaller closer to the base of the jet.
}
\item{
  It takes the plasma longer to lose its energy from TeV down to GeV than from that point onwards.
}
\end{enumerate}

In the end, a~mixture of all of these three options would also be possible. However,~there is also an alternative explanation that we will sketch in the next~subsection.

\subsection{Are the TeV Photons Produced by a Different Mechanism?} \label{sec:mr}

The fact that the \ltm{} phase-offset values for the radio, X-ray, and~GeV emission are perfectly fit by a straight line as a function of $\log(E)$, while the TeV data point is offset from that simple trend by $6\sigma$ {(or even $27\sigma$ with alternative ambiguity resolution for that data point, as~pointed out in Section~\ref{sec:longtermsys})} justifies the consideration of a different or at least modified explanation for the TeV emission. Furthermore, the~fact that the absolute phase-offset of~$\sim$0.5 with respect to the radio means that the \ltm{} pattern of the TeV is in perfect anti-phase with the radio, and~deserves special attention. That the VHE emission is affected by the \ltm{}, i.e.,~the beating between orbit and precession, implies that these photons are produced in a mechanism that is subject to both accretion and periodic changes in the relativistic beaming. One possible explanation could be that during magnetic reconnection events, which can occur both in the accretion disk \citep{Yuan2009} or the jet itself \citep{Petropoulou2016, Sironi2016}, plasmoids are injected into the steady jet and collide with the slower material there. This results in shocks in which the magnetic field lines are compressed into one direction and force the plasmoids to move into a different direction than the steady jet. Consequently, the~Doppler factor would be systematically different for the TeV emission than for the other wavelengths. In~this scenario, a~localization of the TeV emission relative to the other emissions, as~depicted in Figure~\ref{fig:sketch} is not possible any more. It is worth of note that the short-term variability in \lsi{}, observed in the radio \citep{Jaron2018} and X-ray \citep{Noesel2018} emission, mentioned in the beginning of Section~\ref{sec:intro} of this article can be explained in the same scenario of magnetic~reconnection.

\subsection{What Is the Origin of Optical Emission in \lsi{}?} \label{sec:optdisc}

As mentioned in Section~\ref{sec:opt}, there is also observational evidence that the equivalent width of the H$\alpha$ line in \lsi{} is subject to the \ltm{}. {As explained there, w}e excluded these {optical} observations from our {analysis}. In~the context of the physical scenarios presented here, the~origin of the H$\alpha$ line is more complicated. Moreover,~the possibility of the emission line to originate from the circumstellar disk of the Be star, \citet{Fender2009} point out that the accretion disk of X-ray binaries is another source of H$\alpha$ line emission. Furthermore, by~definition EW(H$\alpha$) is a function of both the H$\alpha$ luminosity itself as well as the continuum emission. In addition,~especially in the low/hard state of X-ray binaries, the jet can contribute significantly in the optical band \citep{Russel2006}. The~continuum optical jet emission, in~this scenario, is subject to variable Doppler boosting in the same way as the other wavelengths {emitted from the precessing jet}, and~consequently the resulting \ltm{} of the continuum emission would propagate into the modulation of EW(H$\alpha$). Further observations at optical wavelengths and investigation of the origin of the optical emission {are} needed to clarify these~issues.

{At present the largest void in Figure~\ref{fig:phase} is between radio and X-rays. A~reliable determination of the \ltm{} phase-offset at optical wavelengths would provide data just in the center of this gap. For~this reason alone, optical data would be a natural first choice to be added to a future extension of the analysis presented here. However, before~that, existing databases of optical observations will have to be revisited, and~continuum and line emission will have to be carefully disentangled. In~the scenario suggested in Section~\ref{sec:cool}, the~optical continuum emission is expected from a steady jet originating from a region between those of the radio and X-ray emission. In~this case the \ltm{} phase-offset of the \emph{continuum} optical emission alone would be expected to be $\sim$0.1, interpolating the linear trend that extends from radio to GeV (orange line in Figure~\ref{fig:phase}). Concerning the H$\alpha$ emission line, sources of this line can be \emph{both} the circum-stellar disk of the Be star and the accretion disk, so a variability in this emission line probes variability in both of these regions. Furthermore, in~a feasibility study presented by \citet{Massi2010}, the precessing jet is the result of a precessing accretion disk. Hence, the~H$\alpha$ emission from the accretion disk would be subject to the same intrinsic periodicites, i.e.,~accretion rate modulated by the orbital period~$P_1$, and~variable Doppler boosting modulated by the precession period~$P_2$, resulting in the beating that we identify with the \ltm{}. In~this scenario the \ltm{} phase-offset of this H$\alpha$ emission could be approximately anti-correlated with the TeV emission, if~the TeV emission originates at the base of the steady jet, as~proposed in Section~\ref{sec:cool}. In~this way, optical observations have the potential to sigificantly contribute in emission model discrimination for \lsi{}.}

\section{Conclusions} \label{sec:conclusion}

In this article, we investigated the long-term modulation of the \gammaray{}-loud high-mass X-ray binary \lsi{} at multiple wavelengths. The~periodic modulation of $\sim$4.6~years is consistently detected at radio, X-rays, GeV, and~TeV energies by analysis of the past four decades of observations at these wavelengths. Subject of this article was the investigation of systematic phase-offsets of the modulation pattern between these wavelengths and the explanation in a physical scenario. These are our conclusions:

\begin{enumerate}
\item{
  By re-analyzing archived TeV data \citep{Ahnen2016} we determined that \ltm{} pattern at TeV is offset from the pattern at radio by $0.51 \pm 0.03$. We point out that, while this value fits into the monotonically incrasing trend with the other wavelengths, this also means perfect anti-correlation with the radio emission in terms of the \ltm{} pattern.
}
\item{
  The \ltm{} of \lsi{} is established at radio, X-ray, GeV, and~TeV wavelengths by long-term monitoring of the source. There is a systematic trend of the modulation pattern being increasingly offset from the radio pattern as the energy increases. Emission at higher energy is  lagging emission at lower energy in a strictly monotonically increasing manner (Figure~\ref{fig:phase}). 
}
\item{
  We extended the physical scenario first introduced by \citet{Jaron2018} to X-rays and TeV. In~this scenario, the emission regions are located closer to the base of the jet as the energy increases (Figure~\ref{fig:sketch}). Because~the \ltm{} is the result of interference between orbit and precession, an~earlier peak in the precession profile propagates to a later peak in the \ltm{} pattern with increasing energy, as~observed. We also note that while the radio, X-ray, and~GeV emissions are fitted by a very simple trend, the~TeV emission appears to be significantly offset from that. This is a challenge for this scenario at TeV~energies.
}
\item{
  We consider an alternative scenario in which the TeV photons are produced in shocks resulting from the injection of plasmoids during magnetic reconnection events. It is interesting to note that this would imply a connection between the large phase-offset of the \ltm{} at TeV energies and the hour time-scale phenomenon of quasi periodic oscillations at radio and X-rays. This would explain two phenomena, occurring at different energy bands and on time-scales of different orders of magnitude (hours vs. years) in the same coherent picture. This should be subject of future investigations.
}
\end{enumerate}

Further long-term monitoring of \lsi{} at more wavelengths is needed in order to populate Figure~\ref{fig:phase} with more data points. This would enable a better quantification of the relationship between phase-offset and energy. At~the moment functional fits are complicated by the small number of available data points. In~addition, a~better sampling at X-rays and especially TeV would allow quantitatively studying the precession profile at these wavelengths and to verify whether a shift in this pattern is indeed present. This would also help to disentangle the two scenarios for the TeV emissions and answer the question whether these photons are produced in the steady or transient~jet.

\vspace{6pt} 




\funding{
  This research was funded by the Austrian Science Fund (FWF) [P31625].
}



\dataavailability{
  The very high energy \gammaray{} data analyzed in Section~\ref{sec:dettev} and plotted in Figure~\ref{fig:TeV} are publicly available in \citet{Ahnen2016}.
}

\acknowledgments{\textls[-5]{I thank Mikhail Lisakov from the MPIfR Bonn for carefully reading the manuscript and providing useful comments. Maria Massi provided important information about the multi-wavelength properties of X-ray binaries and \lsi{} in many very fruitful discussions. {The anonymous referees provided constructive comments and very useful information during the review process for this article.}}}

\conflictsofinterest{
  The author declares no conflict of interest.
}


\abbreviations{Abbreviations}{
The following abbreviations are used in this manuscript:\\

\noindent 
\begin{tabular}{@{}ll}
  BAT      & Burst Alert Telescope\\
  EW       & Equivalent Width\\
  INTEGRAL & INTErnational Gamma-Ray Astrophysics Laboratory\\
  LAT      & Large Area Telescope\\
  LTM      & Long-Term Modulation\\
  MAGIC    & Major Atmospheric Gamma Imaging Cherenkov Telescopes\\
  MFF      & Modulated Flux Fraction\\
  OVRO     & Owens Valley Radio Observatory\\
  PCA      & Proportional Counter Array\\
  RXTE     & Rossi X-ray Timing Explorer\\
  VERITAS  & Very Energetic Radiation Imaging Telescope Array System\\
  VHE      & Very High Energy\\
  VLBA     & Very Long Baseline Array\\
  VLBI     & Very Long Baseline Interferometry
\end{tabular}}

\end{paracol}
\reftitle{References}




 \changeurlcolor{black}

%


\end{document}